\documentclass[twocolumn,showpacs,amssymb,preprintnumbers,prc,floatfix]{revtex4-1}
\usepackage{epsfig,dcolumn}
\usepackage{graphicx,bm}
\newcommand{\be}{\begin{equation}}
\newcommand{\ee}{\end{equation}}
\newcommand{\ba}{\begin{eqnarray*}}
\newcommand{\ea}{\end{eqnarray*}}

\begin{document}
\title{The merging of the islands of inversion at N=20 and  N=28}

\author{E.~Caurier,$^1$  F.~Nowacki,$^1$ and  A.~Poves,$^{2,3}$}

\affiliation{$^1$IPHC, IN2P3-CNRS and Universit\'e Louis Pasteur, 
             F-67037 Strasbourg, France\\
	 $^2$Departamento de F\'isica Te\'orica and IFT-UAM/CSIC, Universidad Aut\'onoma de Madrid,  E-28049
	     Madrid, Spain\\ $^3$Isolde (Cern) 1211 Gen\`eve 23, Switzerland}

\begin{abstract}

  The N=20 and N=28  "islands of inversion" are described by large scale shell model
  calculations with an extension of the
  interaction {\sc sdpf-u} that makes it possible to mix
  configurations with  different N$\hbar \omega$ or equivalently with different number
  of particles promoted from the $sd$-shell to the $pf$-shell.
  It allows to connect the classical sd-shell calculations
  below N=18 with the sd(protons)-pf(neutrons) calculations beyond N=24-26,
  for all the isotopes from Oxygen to Argon, using the same interaction.
  For some isotopes this range contains all the nuclei between the proton
  and the neutron drip lines and includes the N=20 and N=28 islands of
  inversion.  We shall pay particular attention to the properties of the states at fix  N$\hbar \omega$ 
  which turn out to be the real protagonists of the physics at N=20. The existence of
  islands of inversion/deformation will be explained as the result of the competition between the
  spherical mean field which favors  the 0$\hbar \omega$ configurations and the nuclear correlations which
  favor the deformed N$\hbar \omega$ configurations.   The Magnesium chain is  exceptional 
  because in it, the N=20 and N=28  "islands of inversion"
  merge, enclosing all the  isotopes between N=19 and N=30.  Indeed, this would be also the case 
  for the Neon and Sodium chains
  if their drip lines would reach N=28.

\end{abstract}


\date{\today}
\maketitle

\section{Introduction}

    At the neutron 
rich edge, the structure of the spherical mean field may be at
variance with the usual one at the stability line. The reason is that,
at the stability line, the T=0 channel of the nucleon-nucleon
interaction has a stronger weight relative to the T=1 channel than it
has when the neutron excess is very large. If the spherical mean field gaps get reduced,
open shell configurations, usually two neutron excitations across the
neutron closure, take advantage of the availability of open shell
protons to build highly correlated states that can  be more bound than the
closed shell configuration.  Then the shell closure is said to have
vanished.   Although it was known since long that the ground state parity 
of $^{11}$Be was at odds with the naive shell model picture \cite{Talmi60},
this fact was overlooked until much later, in connection with the discovery of halo nuclei with $N=8$.
Studies of charge radii, atomic masses and nuclear spectra in the Mg and Na isotopic
chains  did show that a region of deformation exists around $N=20$ below $^{34}$Si.
Key experimental references are gathered in
\cite{32Mg_def-exp,32Mg-coulex,gerda,expN20}.
Since then, a lot a experimental and theoretical work has ensued.  Early mean field calculations suggested
that deformation was responsible for the excess of binding of $^{31}$Na \cite{32Mg_theo_mf}, but at this stage
to get a deformed minimum required the inclusion of "ad hoc" rotational corrections.
In the framework of the shell model, the deformation in the region was soon associated with the dominance of
 two-particle -two-hole $(2p-2h)$ excitations across the $N=20$
shell gap between the normally occupied neutron $d_{3/2}$ orbit and the valence $f_{7/2}$ and $p_{3/2}$ orbits  ~\cite{32Mg_def-theo}. 
These configurations  were dubbed intruders since they do not obey the normal filling of the standard
spherical mean field. More recent shell model works include the MCSM calculations of the Tokyo group \cite{otsu}
and other large scale calculations in the $sd$-$pf$ valence space \cite{sm**2}. Beyond mean field calculations have
also been used in the description of the region with diverse degrees of success \cite{bmf}.

The interaction  {\sc sdpf-u}  \cite{now09} that we proposed some time ago, was aimed to the description of 
the very neutron rich nuclei around  N=28  in a  0$\hbar \omega$ space, with valence protons
in the $sd$-shell and valence neutrons in the $pf$-shell. Therefore, it is applicable
only to nuclei with 8$\le$Z$\le$20 and 20$\le$Z$\le$40 and does not describe
intruder states. The main asset of {\sc sdpf-u} was the description
of the vanishing of the N=28 shell closure below $^{48}$Ca, most notably
in $^{42}$Si \cite{bas07}  (a result which is now fully verified \cite{tak12}, 
but which produced initially some heated debates \cite{Fri05}).
$^{42}$Si  was predicted to be oblate deformed and $^{40}$Mg prolate
deformed, exhibiting perhaps a neutron halo.  Since its publication, it has
been frequently used and shown to give an excellent description of this region
of  very neutron rich nuclei \cite{n28exp}.  Very recently, these calculations have
been repeated in the same valence space with a somewhat different effective interaction, 
getting (as could be expected) very similar results  \cite{42si:tensor}. As the $sd$ part of  {\sc sdpf-u} is just the  {\sc usd} 
interaction \cite{USD} and its $pf$  part a variant of {\sc kb3} previous to {\sc kb3g} , it is appealing to complete
 {\sc sdpf-u} with the $sd$-$pf$ off-diagonal
matrix elements and to retune the  $sd$-$pf$  cross shell monopoles in such a way
that the  {\sc sdpf-u}  results at  0$\hbar \omega$ are mostly preserved and the $sd$-$pf$ gaps are
in accord with the experiment. This process results in 
the  {\sc sdpf-u-mix}  interaction. More details  are given in Appendix A.
The calculations are carried out using the codes {\sc antoine} and {\sc  nathan} \cite{Cau05}
and reach basis dimensions  of O(10$^{10}$). In  a  (very) loose sense one can pretend that this interaction covers the sector of the Segr\'e
chart  8$\le$Z,N$\le$40. In this article we shall concentrate in the physics of the N=20
"island of inversion" and its merging in some cases with the neighboring N=28 one.

\section{The physics at fixed  N$\hbar \omega$}

   What is the driving force behind the abrupt changes leading to the appearance of  these
  "islands of inversion"?  What makes these intruder states special? That they need to 
  be highly correlated in order to compensate for the energy loss associated to the breaking
  of the normal filling of the spherical mean field. Obviously, small gaps are easier to overcome,
  thus a reduction of the neutron magic gaps  at the very neutron rich edge is good news for the intruders.
   The mechanisms need not to be the same in the different regions. For instance in
    $^{11}$Li the intruder is mostly pairing boosted while in  $^{11}$Be the quadrupole interaction
    is more important. In the other three neutron rich regions, N=20, N=28 and N=40, the quadrupole interaction 
    is the main player. Let us concentrate in the   N=20 case. Compared to the configurations with
    closed N=20, the intruders (np-nh) have neutrons in open $sd$ and  $pf$-shell orbits and in some cases
    protons in open $sd$- shell orbits.  This favors the efficient build up of correlations by the 
     neutron-proton quadrupole interaction
    when the open orbits are the appropriate ones. And whose are these is dictated 
    by the different variants of SU(3). For instance,  when valence neutrons or protons  occupy quasi-degenerate orbits
     with $j_r - j_s$=2 and $l_r - l_s$=2 the coupling scheme is Quasi-SU(3) \cite{quasi},  if they are in
    quasi-spin doublets  the regime is that of Pseudo-SU(3) \cite{pseudo} . In the limit of vanishing spin-orbit splitting,
    all the orbits in a harmonic oscillator shell form an Elliot's SU3 multiplet \cite{su3}.  To get large coherence the neutrons 
    and the protons must pertain to one or another of these coupling schemes.   For example,  in the case of the N=20
    intruders, the neutrons in the orbits  0f$_{7/2}$ and 1p$_{3/2}$,  and the protons in  
    0d$_{5/2}$ and 1s$_{1/2}$   are in the Quasi-SU(3) regime  and the neutrons in 0d$_{3/2}$ and 1s$_{1/2}$
     in Pseudo-SU3.

 \begin{figure}
\begin{center}
\includegraphics[width=1.1\columnwidth,angle=0]{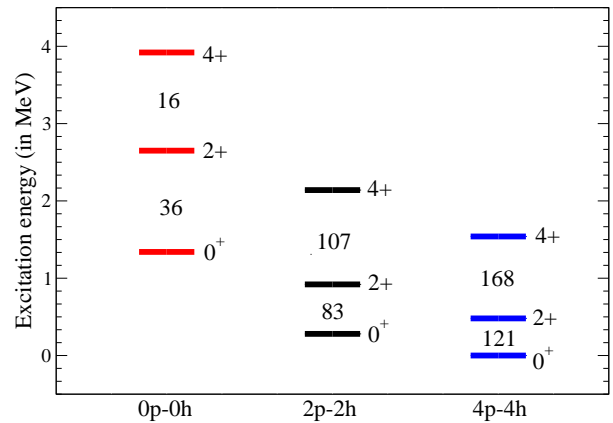}
\end{center}
\caption{(Color online)  The low energy spectra and B(E2)'s (in e$^2$fm$^4$) of the 0p-0h, 2p-2h and 4p-4h configurations in $^{32}$Mg. \label{spec}}
\end{figure}  

   Let's make these statement quantitative in a few selected cases. In this section all the calculations are performed at fixed N$\hbar \omega$.
  We only allow neutron jumps from the $sd$ to the $pf$-shell without any other truncation. We have verified that the effect of the proton excitations 
  to the $pf$-shell is negligible. We take care of the (small) center of mass  contamination by adding to the effective interaction
  the center of mass hamiltonian (with $\hbar \omega$=A).  The expectation value of the center of mass hamiltonian in the
  physical states is always below  0.001A. The  results for  the low
  energy levels of  $^{32}$Mg are presented in Figure~\ref{spec}. We can follow the evolution from the semimagic
  0p-0h result, with a high excited 2$^+$ and a low B(E2) to a rotational-like 2p-2h whose B(E2) corresponds to
  $\beta$=0.4/0.5 and finally to a perfect rigid rotor 4p-4h with \mbox{E(4$^+$)/ E(2$^+$)=3.2} and a very large B(E2) that  corresponds to
  a super-deformed structure. Most important for our aims is that the gains in energy due to the correlations 
 --defined as the difference between the energy which comes out of the diagonalization  and the energy of the
 lowest 0$+$ state of seniority zero in the corresponding space--  are very different in the 0p-0h, 2p-2h and 4p-4h
  spaces; 1.5~MeV, 12.5~MeV and 21~MeV respectively. 
  These huge correlation energies may  eventually overcome
  the spherical mean field gaps. In fact this is the case in $^{32}$Mg. With  {\sc sdpf-u-mix}   
  the lowest 4p-4h 0$^+$ state is  about  250~keV below 
  the lowest  0$^+$  of the 2p-2h space and 1.2~MeV below the 0$^+$  of the 0p-0h configuration.   
  This  near degeneracy of the 2p-2h and 4p-4h bandheads is not a spurious manifestation of 
  our spherical mean field not producing the right $sd-pf$ gap, but 
   due to the fact that the  the energy gain per particle promoted to the
  $pf$-shell, is the same for both configurations.  We want to stress again the fact
  that, in favorable circumstances like these, the gain in correlation energy of the intruders can beat the spherical mean field. In fact,
  in the laboratory frame, this is the microscopic mechanism responsible for the shape transitions from spherical
   to deformed nuclei \cite{quasi}.
  The lowest negative parity state of 1p-1h nature is a 3$^-$, 4~MeV above the 2p-2h 0$^+$ and the lowest 3p-3h state
  is  a   2$^-$, 2.5~MeV above the 2p-2h 0$^+$. Their respective energy gains are  5~MeV and 16~MeV and the 
  underlying structures correspond to the K=3$^-$ and   K=2$^-$ band-heads as expected from the Nilsson diagrams for
  $\beta$=0.15 and  $\beta$=0.4.

   The 4p-4h state of  of $^{32}$Mg has an academic interest in itself even if  the states belonging to its rotational band do
    not manifest themselves openly in the low energy spectrum, 
    (as do their cousins in the superdeformed bands of  $^{36}$Ar and  $^{40}$Ca  \cite{sve00,ide01,cmnp:07})
     because of its strong mixing with the
   0p-0h and 2p-2h spherical and deformed states. It may well happen that they could become yrast at some higher spin,
   but  the threshold for neutron emission is not very high, and the experiments to find them are probably hopeless. 
   In fact, one can understand semi-quantitatively why this configuration can produce such  superdeformed structure in the context of  Elliott's
   SU3 and it variants.  Let's assume that the  four $pf$-shell neutrons be in Quasi-SU3  and the four neutron holes in $sd$ in  Pseudo-SU3;
   in this case the neutrons will contribute with 24b$^2$ (times the effective charge) (b is the harmonic oscillator length parameter)
   to the intrinsic quadrupole moment. If we go to the SU3 limit in the $pf$-shell sector this number increases to  
   26b$^2$. The value from the shell model calculation is 24.7b$^2$. For the protons, the Quasi-SU3 limit gives 11b$^2$
   against 9.7b$^2$ (times the effective charge) of the shell model calculation.   With effective charges 0.46 and 1.31  
     for neutrons and protons, taken from
   the work of Dufour and Zuker \cite{dz},
these values
   lead to  $\beta$=0.6/0.7 depending of the definition of $\beta$.
   
   It follows from the above discussion that the configuration with four neutrons in the $pf$-shell and two neutron holes in
   the $sd$-shell maximizes the quadrupole moment and, {\it a fortiori} the quadrupole correlation energy. Therefore one
   should expect the 2p-2h configuration to be also dominant in $^{34}$Mg. On the contrary one expects the \mbox{0p-0h} 
   one to begin taking
    over in   $^{36}$Mg.
   This would establish the limit of the N=20 "island of inversion". However, as we shall see in the next section, the 
   very large depopulation of the 0f$_{7/2}$ orbit in   $^{36}$Mg  indicates that  before leaving the  N=20  "island of inversion"
    we  enter the  N=28  "island of inversion", 
   and that both islands are actually  merged in a single one. These arguments apply as well to the 3p-3h excitations in the N=21
   isotopes, that we expect very low in energy.

\begin{figure}
\begin{center}
\includegraphics[width=1.1\columnwidth,angle=0]{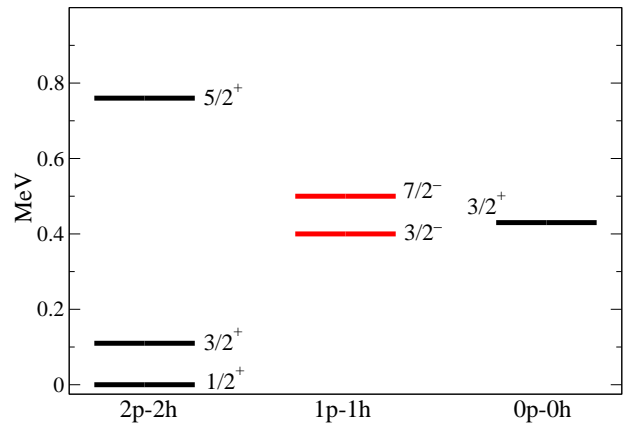}
\end{center}
\caption{(Color online) The low energy spectra of the 0p-0h, 1p-1h and 2p-2h configurations in $^{31}$Mg. Energies relative to the
 2p-2h  $\frac{1}{2}^+$ state\label{mg31tf}}
\end{figure}

    In $^{31}$Mg the configurations
    0p-0h, 1p-1h and 2p-2h are nearly degenerate. The lowest one is the 2p-2h which looks like a K=$\frac{1}{2}^+$ band, with an
    excited $\frac{3}{2}^+$ at  $\sim$100~keV, in agreement
    with the experimental findings of refs. \cite{gerda}. The energy gain of the band is 14.5~MeV. The lowest \mbox{0p-0h} state, a    $\frac{3}{2}^+$,
    gains just 3.5 MeV and is 400~keV less bound than the 2p-2h   $\frac{1}{2}^+$. The lowest \mbox{1p-1h} negative parity state,  a $\frac{3}{2}^-$,
    gains 8.5~MeV and is 400~keV above the  {2p-2h}   $\frac{1}{2}^+$. These results are gathered in Figure~\ref{mg31tf}. 
    The E2 and M1 transition probabilities of the 2p-2h band compare well with the recent experimental values from 
    ref.~\cite{Seidlitz}:
    
     \mbox{B(M1)($\frac{5}{2}^+  \rightarrow \frac{3}{2}^+ $)=0.1-0.5~$\mu_N^2$}; (th.  0.35~$\mu_N^2$) 
       
    \mbox{B(M1)($\frac{3}{2}^+  \rightarrow \frac{1}{2}^+ $)=0.019(4)~$\mu_N^2$}; (th. 0.03~$\mu_N^2$) 
      
    \mbox{B(E2)($\frac{5}{2}^+  \rightarrow \frac{1}{2}^+ $)=61(7)~e$^2$fm$^4$}, (th.  84~e$^2$fm$^4$)  
     
    The 
    magnetic moment of the    $\frac{1}{2}^+$  (using bare g-factors) is  --0.85~$\mu_N$ very close to the experimental value
   \mbox{ --0.88355(15)~$\mu_N$}, thus we can expect that its 2p-2h character is rather pure, the more so in view of the absence
    of nearby $\frac{1}{2}^+$ states to mix with.   The intrinsic quadrupole moment of the ground state band is the typical in this region,
     Q$_0$$\approx$70~efm$^2$  
      
\begin{figure}[h]
\begin{center}
\includegraphics[width=1.1\columnwidth,angle=0]{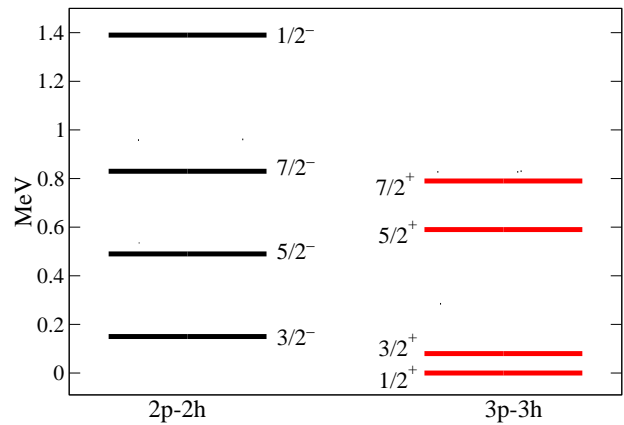}
\end{center}
\caption{(Color online) The low energy spectra 
of the  2p-2h  and 3p-3h configurations in $^{33}$Mg. Energies relative to the
 3p-3h  $\frac{1}{2}^+$ state\label{mg33tf}}
\end{figure}

  In $^{33}$Mg the lowest state at fixed configuration is the 3p-3h  $\frac{1}{2}^+$, head of a  K=$\frac{1}{2}^+$ band.
  At 150~keV  appears the  \mbox{2p-2h} $\frac{3}{2}^-$, head of a K=$\frac{3}{2}^-$ band. 
  The 0p-0h and \mbox{1p-1h} states lie  more  than 1.5~MeV higher.
  These results are gathered in Figure~\ref{mg33tf}. Both  structures are highly collective, with B(E2)'s in excess of 100~e$^2$fm$^4$. 
  In particular the
  K=$\frac{1}{2}^+$ 3p-3h  band can be viewed as the addition of two neutrons  to the ground state band of  $^{31}$Mg. It turns out that both 
  band heads, in spite of their different spin and parity have negative magnetic moments  (--0.49~$\mu_N$ for the 2p-2h and --0.87 for the 3p-3h).
  Contrary to the assumption of ref.~\cite{Tripathi}, the magnetic moment of the   $\frac{3}{2}^+$  3p-3h state is positive (+0.62~$\mu_N$).  
  The results of the fully mixed calculation will be discussed in Section~VI.

 Similar analysis can be carried out for the all the remaining isotopes. We want to underline here two important points: i)
 The configurations at fixed np-nh contain much of the relevant physics,  and ii) In the cases
 in which configurations with
 different particle hole structures, and hence with very different amounts of energy gains due to the correlations, compete, as  in some
 N=19 and N=21 isotopes,  the final balance
 between monopole energy losses and correlation gains is very delicate and  the difficulty in accounting for
 experimental energy splittings which may be smaller than 100~keV, extreme.
 
%

\section{Spherical mean field versus Correlation energies: The mechanism of configuration inversion}

\begin{figure}
 \begin{center}
    \includegraphics[width=1.0\columnwidth]{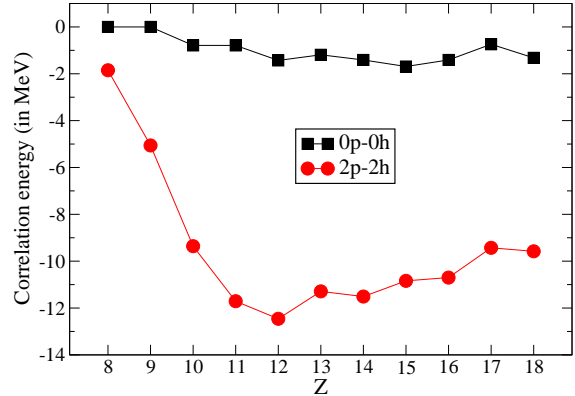}
  \end{center}
\caption{(Color online) Correlation energies of the 0p-0h (squares) and 2p-2h (circles) configurations at N=20 \label{fig:n20corr_odd}}
\end{figure}

 As we have already anticipated, the "islands of inversion" occur when  a group of adjacent nuclei have
 their ground states dominated by intruder configurations.  We develop now the case of the  N=20 isotopes.  We have 
 plotted in Fig.~\ref{fig:n20corr_odd} the correlation energies of the lowest  states of the 0p-0h and 2p-2h 
 configurations.  As the uncorrelated energy we take in each case the lowest diagonal energy (expectation value of the hamiltonian)
 in a basis of states coupled to good J  and with well
 defined generalized seniority. Because of this choice  we incorporate in fact some diagonal pairing energy in our
 uncorrelated reference, but this is irrelevant for our purpose.  As expected for  semi-magic nuclei, these correlation energies
 are small and roughly constant for the 0p-0h configuration. On the contrary, for the 2p-2h intruders, they can be very large and 
 have a  rapid variation with Z. The largest values occur at mid proton shell, when the quadrupole collectivity reaches its maximum.

\begin{figure}
 \begin{center}
    \includegraphics[width=1.0\columnwidth]{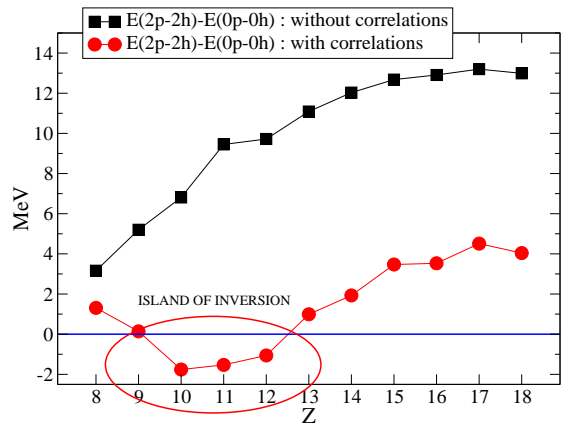}
  \end{center}
\caption{(Color online)  The gap between the 0p-0h and the 2p-2h configurations at N=20. Without correlations (squares) and
including correlations (circles). Nuclei close or below the zero line are candidates to belong to the island of inversion \label{fig:n20odd_gap}}
\end{figure}

In Fig.~\ref{fig:n20odd_gap}  we present the differences in energy between the lowest 0p-0h state with well defined J$^{\pi}$ and the
 lowest 2p-2h state {\bf without correlations}. It is seen that in all cases the normal filling gives the lowest 
 energy, although between Z=8 and Z=14 there is an almost linear increase from 3~MeV to 12~MeV while from there on the curve is
 much flatter. This reflect the reduction of the $sd$-$pf$ gap as we approach the neutron drip line. When we take fully into account
 the correlations the situation changes dramatically as reflected in the lower curve of the figure.  The balance between the 
 correlation gains and the monopole losses of energy defines the borders of the island of inversion at N=20 in  $^{29}$F and  
 $^{33}$Al. Clearly,  $^{30}$Ne, $^{31}$Na, and $^{32}$Mg  are {\it bona fide} members of the club.  Equivalent graphs can be drawn
 for the other isotonic chains. Roughly speaking  the situation is very similar  for the N=19, N=21 and N=22 isotonic chains.
 It is probably not worth to go much more beyond this qualitative definition of the somewhat fuzzy shores of the "island of inversion"
 around N=20 because the predictions obtained in the analysis at fixed configuration may sometimes change when the full mixing 
 is taken into account, the more so for the nuclei near to the borders. We will be more precise in the chapter dealing with the full scale results
 of our calculations.

%


\section{From N=Z to N=32 in the  Mg, Ne, and Si  isotopes}

\begin{figure}[h]
\begin{center}
\includegraphics[width=1.0\columnwidth,angle=0]{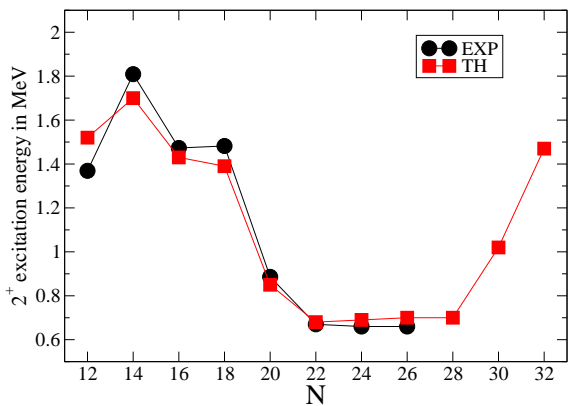}
\end{center}
\caption{(Color online) Excitation energies of the first 2$^+$ state in the Magnesium isotopes; Results of the 
 calculations with the  {\sc sdpf-u-mix}  interaction in the valence space of the sd-shell for the protons and  the sd-pf-shells
 for the neutrons, compared with the available experimental data.   \label{fig:mg20}}
\end{figure}

   In Fig. \ref{fig:mg20} we compare the experimental 2$^+$ excitation energies of the 
  even Mg isotopes, starting at N=Z,  with the shell model calculations with the  {\sc sdpf-u-mix} interaction.  Up to N=16 the
  results should not differ much from to the ones
  produced by the  {\sc usd} interactions \cite{USD}.  Beyond N=16 the calculations include (if necessary for
  convergence) up to 6p-6h excitations
  from the $sd$-shell to the full $pf$.  The agreement is excellent and covers all the span of isotopes 
  from $^{20}$Mg  (which should be close to the proton drip line)  to the neutron drip line. Notice the disappearance of the semi-magic closures at
  N=20 and N=28 and the presence of a large region of deformation which connects the two islands
  of inversion, previously though to be split  apart.    The agreements is really superb. Beyond N=24 the effect of the core 
  excitations is perturbative and produces a small   expansion of the
  spectra which improves slightly the agreement with the experimental data obtained in the 0$\hbar \omega$
  calculations. The merging of the N=20 and N=28 "islands of inversion" is evident.
  
\begin{figure}
\begin{center}
\includegraphics[width=1.0\columnwidth,angle=0]{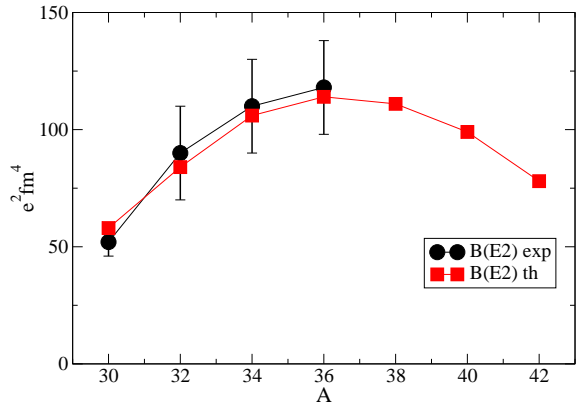}
\end{center}
\caption{(Color online) B(E2)'s of the Magnesium isotopes compared to the experimental results \label{fig:mg_be2}}
\end{figure}

  In Fig.~\ref{fig:mg_be2} we compare the B(E2)'s
  in the transition region with  the experimental data including some  unpublished results from Riken~\cite{riken.mg.be2}. 
  We use  effective charges 1.35 and 0.35 for protons and neutrons respectively,  which are fully
  compatible with a recent fit to the sd-shell nuclei with the interaction {\sc usd-a} \cite{usda} and with the results 
  obtained by Dufour and Zuker in ref.~ \cite{dz}.  We take $\hbar \omega = 45 A^{-1/3} - 25 A^{-2/3}$.
  The agreement is very good as well.

\begin{figure}
\begin{center}
\includegraphics[width=1.0\columnwidth,angle=0]{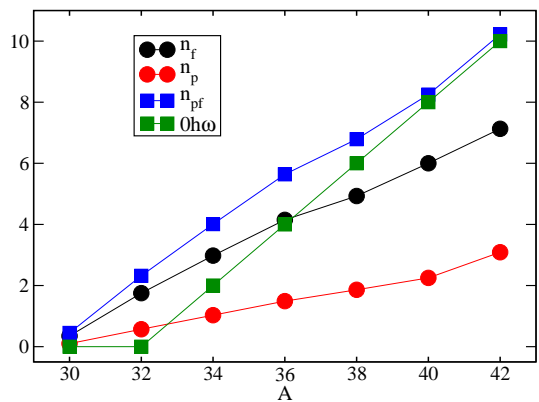}
\end{center}
\caption{(Color online) Occupation numbers of the $pf$-shell orbits in the Magnesium isotopes;  $f$ orbits (black),
 $p$ orbits (red), $pf$-shell (blue) and normal filling (green)  \label{fig:mg_occ}}
\end{figure}

 In Figure~\ref{fig:mg_occ} we have gathered the  occupancies of the $pf$-shell  orbits  in the even-even Magnesium isotopes. We have
 aggregated the values of the $f$ and $p$ orbits for simplicity. The reference numbers for the total  $pf$-shell
 occupancies are those labeled 0$\hbar \omega$ in the figure. The $pf$-shell has more than two neutrons in excess at N=20
 and N=22. At N=24 the excess is of about one  neutron, and beyond that, the core excitations are much damped.
 
 What is more interesting is that when the $sd$-shell core excitations become small, the occupancy of the $p$-orbits
 (mainly 1p$_{3/2}$) keeps increasing so that in N=26 and N=28 about two neutrons are in 1p$_{3/2}$, whereas the
 expected occupancy if N=28 were a strong closure would have been zero. In this sense we can speak of the
 merging of the "islands of inversion" at N=20 and N=28 in the Magnesium isotopes. Notice also that a large
 occupancy of the $p$ orbits favors the appearance of a neutron halo  when the neutron separation energy 
 becomes close to zero as it might be the case in $^{37,39}$Mg  and $^{40}$Mg. Our occupancies for  the $pf$-shell  orbits in
 $^{32}$Mg agree with the experimental results of ref.~\cite{Terry}.

\begin{figure}[h]
\begin{center}
\includegraphics[width=1.0\columnwidth,angle=0]{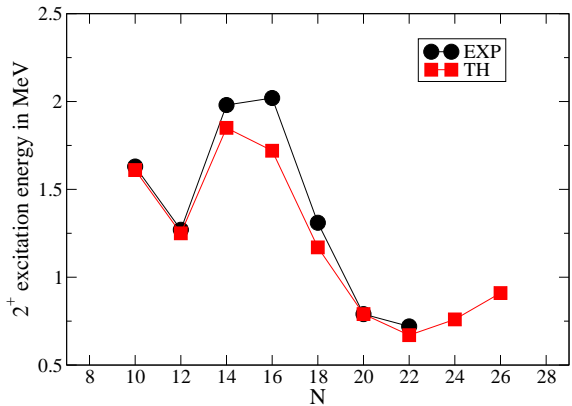}
\end{center}
\caption{(Color online) Excitation energies of the first 2$^+$ states in the Neon isotopes (see caption of  Fig. \ref{fig:mg20}).  \label{fig:ne20}}
\end{figure}

\begin{figure}
\begin{center}
\includegraphics[width=1.0\columnwidth,angle=0]{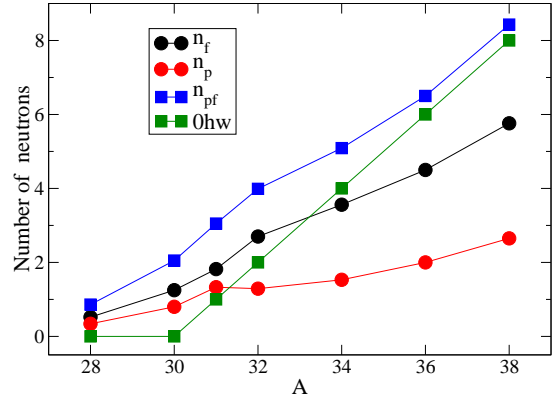}
\end{center}
\caption{(Color online) Occupation numbers of the $pf$-shell orbits in the Neon isotopes;  $f$ orbits (black),
 $p$ orbits (red), $pf$-shell (blue) and normal filling (green) \label{fig:ne_occ}}
\end{figure}

  The results  for the Neon isotopes  (Fig. \ref{fig:ne20})  are very similar to the Magnesiums, although in this case the 
  N=28 isotope $^{38}$Ne is most probably beyond the neutron drip line.  The  2$^+$   excitation energy of  $^{32}$Ne
  is from ref.~\cite{door_ne}.  In Fig. \ref{fig:ne_occ}  we have  collected
  the occupancies of the $f$ and $p$ orbits in the isotopic chain as a function of  the neutron numbers. The behavior
  is very similar to that in the Magnesium chain, except that the $p$ orbits are even more occupied. We have added the
  numbers for $^{31}$Ne, because some recent experimental  data \cite{take} suggest that it could develop a neutron halo. 
  Indeed, ours results are consistent with this hypothesis because  the 1p$_{3/2}$ orbit has on average more 
  than one neutron.

\begin{figure}
\begin{center}
\includegraphics[width=1.0\columnwidth,angle=0]{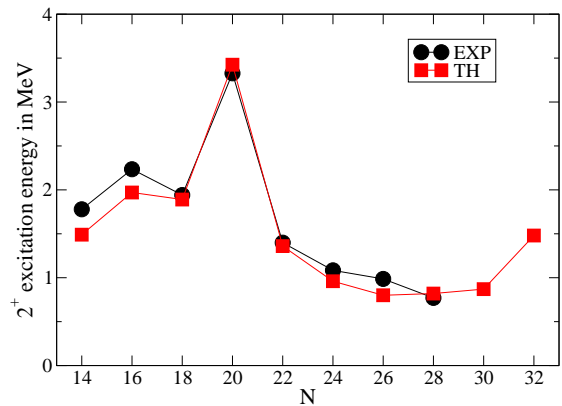}
\end{center}
\caption{(Color online) Excitation energies of the first 2$^+$ states in the Silicon isotopes 
 (see caption of  Fig. \ref{fig:mg20}). 
 \label{fig:si20}}
\end{figure}

    In Fig. \ref{fig:si20}
  we show the results for the Silicon isotopes  (notice the very different energy scale).   At variance with the Magnesium case, we observe a
  majestic peak at N=20, fingerprint of the double magic nature of $^{34}$Si which we will discuss in more detail later, and, as in the Neon and Magnesium cases, no trace
  of the N=28 shell closure is seen, in agreement with the findings of recent experiments at Ganil \cite{bas07} and Riken \cite{tak12}.

\section{Landing at  the island of inversion;  $^{30}$Mg $\rightarrow$$^{32}$Mg   and $^{34}$Si $\rightarrow$$^{32}$Mg}

There are two courses to
 land at the "island of inversion" by the $^{32}$Mg shore : through the isotopic and the isotonic chains. 
 Both are of paramount importance for the understanding of the rich variety of structural changes which
  take place in the region. Adding two neutrons to $^{30}$Mg provokes the inversion of the normal and 
  intruder configurations which are shifted by nearly 3~MeV in $^{32}$Mg. In the isotonic course  the transition
 is even more abrupt as has been recently shown in a GANIL experiment \cite{rot12}:  by removing 
  two protons from $^{34}$Si, the intruder (deformed) state is  shifted down by about 4~MeV 
  with respect to the spherical one to become the ground state of  $^{32}$Mg. 
  
\begin{figure}
\begin{center}
\includegraphics[width=1.1\columnwidth,angle=0]{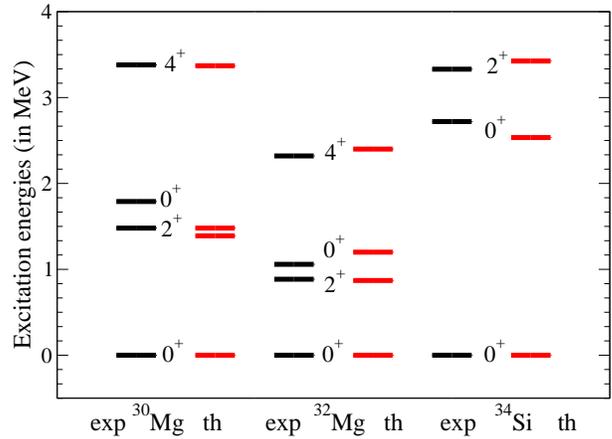}
\end{center}
\caption{(Color online)  Comparison between experiment and theory for the most important low lying states in 
$^{30}$Mg,  $^{32}$Mg   and $^{34}$Si  \label{30.32.34}}
\end{figure}

\indent We compare the experimental data with the shell model results  in Figure~\ref{30.32.34}.  The calculations include configurations with up to 6 neutrons on the $pf$-shell.  In $^{30}$Mg and $^{34}$Si the ground states are
dominantly ($>$80\%) 0p-0h and the first excited 0$^+$'s dominantly 2p-2h.   They differ in the structure of the lowest 
2$^+$ which is 0p-0h in 
$^{30}$Mg and 2p-2h in $^{34}$Si. More 
details on this last nucleus can be found in reference \cite{rot12}.  

The structure of the 0$^+$ states in $^{32}$Mg  is extremely 
    singular; the ground state has  9\% 0p-0h, 54\% 2p-2h, 35\% 4p-4h and 1\% 6p-6h, while the excited  0$^+$ has   
  33\% 0p-0h, 12\% 2p-2h, 54\% 4p-4h and 1\% 6p-6h.  The 2$^+$ has a structure similar to the ground state. As shown in 
  Fig.~\ref{fig:mg_be2} its B(E2) agrees with the experimental result. In addition, the calculated spectroscopic quadrupole moments
  of the  2$^+$ and  4$^+$ states and the B(E2)'s in the yrast band are compatible with a single intrinsic state with 
  Q$_0$$\approx$65~e~fm$^2$. The MCSM calculations of ref.~\cite{otsu}, which
  only include the 0f$_{7/2}$ and  1p$_{3/2}$ orbits of the $pf$-shell, give results similar to ours except for the excited 0$^+$ which is too high by
  almost 2~MeV. 
  Similarly, in $^{34-40}$Mg the calculated E2 properties are compatible with   Q$_0$$\approx$70~e~fm$^2$, which is another fingerprint of the merging of the N=20 and N=28 islands of inversion/deformation.

  Since the early beta decay experiments at Isolde \cite{klotz} it is known that in $^{32}$Mg there are many states, mostly of
 negative parity, above the 4$^+$. They have been explored more recently  via the   $^{32}$Na beta decay  \cite{Mattoon,Tripathi} or
 in (p,p') experiments \cite{Takeuchi-2}.  Ref.~\cite{Tripathi} presents also the MCSM predictions for the negative parity states
 fed in the beta decay. The experimental level at 2.551~MeV is most probably the second 2$^+$. MCSM puts it at  3~MeV whereas
 we get it at nearly the same energy than the 4$^+$.  According to these references, the lowest experimental negative parity state
 would appear at 2.858~MeV. The calculated negative parity states are  1$^-$ at 3.0~MeV, 2$^-$ at 3.1~MeV,  3$^-$ at 3.4~MeV.
 4$^-$ at 3.9~MeV, 0$^-$ at 4.0~MeV,  5$^-$ at 4.2~MeV. They are 
  mostly of 3p-3h nature. 
  The lowest negative parity states  in  the MCSM description are of 3p-3h nature as well, and start  at 3.8~MeV
 with four close packed states  (2$^-$, 1$^-$, 2$^-$, 3$^-$)  followed by a doublet (4$^-$, 5$^-$) at about 4.5~MeV.

\section{Miscellaneous results}

\noindent
{\bf $^{31}$Mg  and $^{33}$Mg.}
\medskip

 The N=19 and 21 isotonic chains are very complex, because of the near degeneracy of configurations with different particle hole structure,
 as discussed in Section II. In $^{33}$Mg the situation is especially critical. We have not tried to fine-tune the interaction to  improve our
 results which amount to have the  $\frac{1}{2}^+$, $\frac{3}{2}^+$ and $\frac{3}{2}^-$ states degenerated (see Table~\ref{tab:mg31-33}). 
 The magnetic moment of the $\frac{1}{2}^+$ ground state  has been measured as  --0.88355(15)~$\mu_N$ \cite{gerda}.  Our predictions 
 show a very strong dependence on the choice of the gyromagnetic factors and marginally agree with the experimental value.

In $^{33}$Mg the fully mixed calculation produces a  $\frac{3}{2}^-$ ground state with the    $\frac{1}{2}^+$
  state  just 40~keV higher.  The magnetic moment of the   $\frac{3}{2}^-$, --0.54$\mu_N$(b); --0.49$\mu_N$(e),
  is short from the experimental
  value,  --0.7456(5)~$\mu_N$.  
  on the nature of the ground state of $^{33}$Mg. If  the J=3/2 assignment is firm, thus positive parity is excluded by the sign of
  the magnetic moment. Notice however that our results locate the positive parity states almost degenerated with the ground state.
  Indeed they  should  not be taken as the theoretical counterparts of the experimental positive parity states at 546~keV and 705~keV.
  The magnetic moment of the 4p-4h $\frac{3}{2}^-$ state is  --1.67~$\mu_N$(b) (--1.36~$\mu_N$(e)),  therefore,  a  somewhat larger  mixing
  of 4p-4h components than that given by our calculation, may  line up
  the theoretical value  with the experimental one.

%
\begin{table}[h]
\caption{\label{tab:mg31-33} Excitation energies (in MeV) and magnetic moments (in $\mu_N$) for the low lying states of
$^{31}$Mg  and $^{33}$Mg  with (b)are and (e)ffective g-factors. }
\begin{tabular*}{\linewidth}{@{\extracolsep{\fill}}|cccccc|}
\hline  
$^{31}$Mg & J$^{\pi}$ &E(exp)&  E(th) &  $\mu$(th) (b) &  $\mu$(th) (e) \\ 
\hline
 & $\frac{1}{2}^+$  &  0.0  &  0.04 &   --0.93   & --0.65\\ 
  & $\frac{3}{2}^+$  &  0.05  &  0.04 &   +1.13  & +0.81 \\ 
   & $\frac{3}{2}^-$  &  0.221  &  0.0 &   --1.24    & --1.07\\ 
\hline 
 $^{33}$Mg & J$^{\pi}$& E(exp)&  E(th) &  $\mu$(th) (b) &  $\mu$(th) (e)\\ 
\hline 
 & $\frac{3}{2}^-$  & 0.0   &  0.0 &  --0.54    & --0.49 \\  
 & $\frac{5}{2}^-$  & 0.484   &  0.33 &  --0.07  & --0.09  \\  
 & $\frac{1}{2}^+$  &    & 0.04  & --0.93    & --0.69 \\  
& $\frac{3}{2}^+$  &    &  0.12 &   +0.69  & +0.44\\  
 \hline 
\end{tabular*}
\end{table}

\medskip
\noindent
{\bf  $^{31}$Na  and $^{33}$Na.}
\medskip

\begin{figure}
\begin{center}
\includegraphics[width=1.0\columnwidth,angle=0]{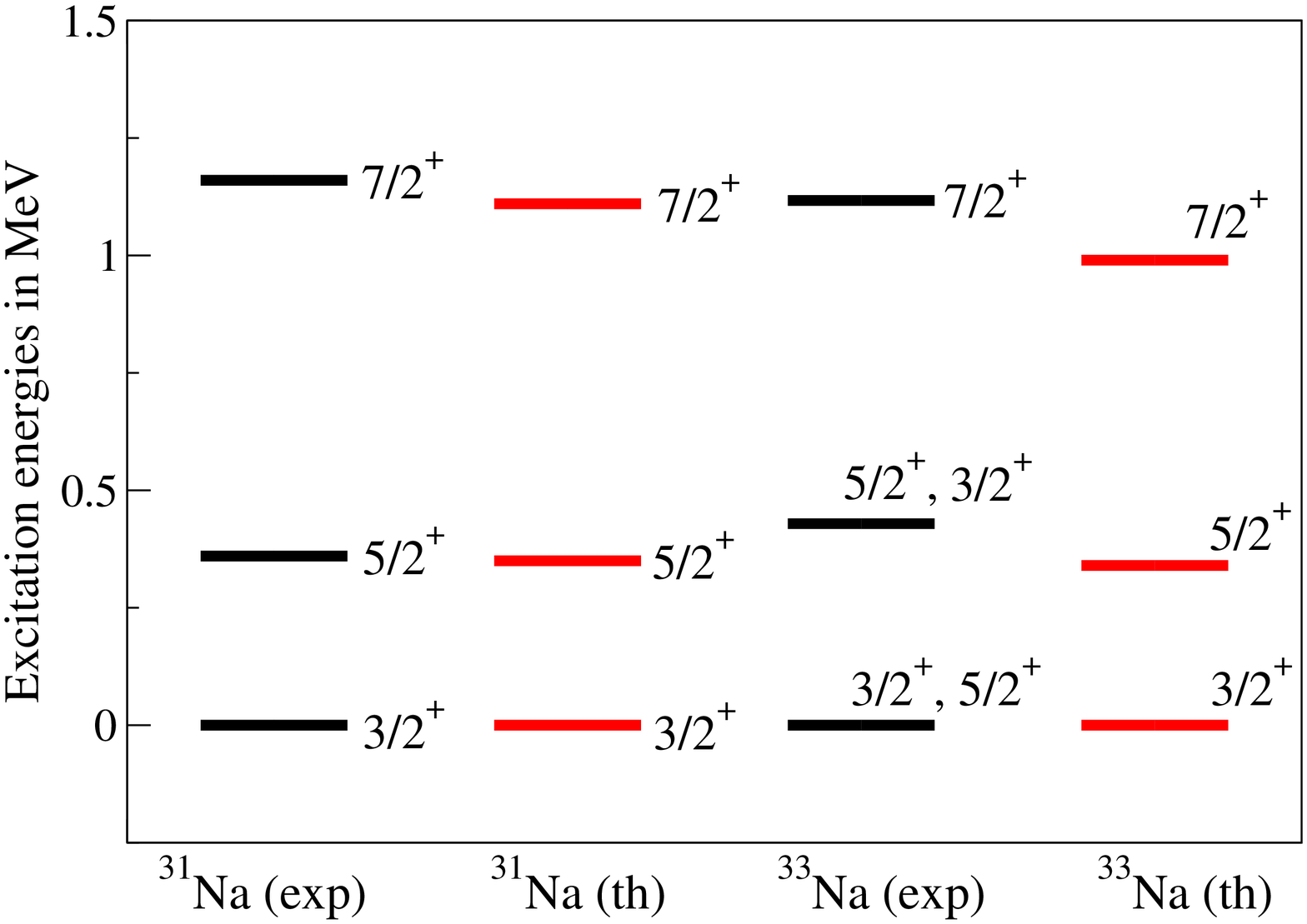}
\end{center}
\caption{(Color online) Comparison of the calculated spectra with the available experimental
data in  $^{31}$Na and $^{33}$Na. \label{na31-33}}
\end{figure}

$^{31}$Na was for many years the protagonist of the N=20 saga, even if only the properties of its ground state were
know (spin-parity, magnetic moment, isotope shift, binding energy).  Although other nuclei have taken the relay
nowadays, it still deserves attention. We have gathered the available  experimental information in   
Figure \ref{na31-33} .  The newest data  \cite{door_na,gade_na33} consists in the excitation energies of two members of the
K=$\frac{3}{2}^+$ ground state rotational band and the B(E2) of the lowest in-band decay in $^{31}$Na, and the
excitation energies of two levels in $^{33}$Na.  Notice the very nice agreement of the calculation and the data which extends to
the ground state magnetic moment (2.298~$\mu_N$ (exp) vs. the calculated 2.26~$\mu_N$ (b) or 1.96~$\mu_N$ (e)). 
As in the  $^{32}$Mg case the calculated E2 properties of $^{31}$Na are compatible with an intrinsic state with Q$_0$~$\sim$65~e~fm$^2$.
We have also plotted the results for  $^{33}$Na in which the behavior of the calculated excitation energies is closer
to J(J+1) than in the previous case, and  Q$_0$~$\sim$~72~e~fm$^2$. The comparison of the new data
with the calculated values is quite good and supports strongly a $\frac{3}{2}^+$ ground state.

\medskip
\noindent
{\bf  $^{30}$Ne,  $^{31}$Ne and $^{32}$Ne.}
\medskip

 To complete this spectroscopic tour, we visit the Neon chain.  Our results for $^{30}$Ne; 
2$^+$ at  0.79~MeV and  4$^+$ at 2.14~MeV compare very well with the experimental values, 0.792~MeV and 2.235~MeV
respectively. The  B(E2),  (2$^+$$\rightarrow$0$^+$)  is predicted at 66~e$^2$~fm$^4$ compared with the measured value
90(54)~e$^2$~fm$^4$. Again a single intrinsic state with  Q$_0$$\approx$60~e~fm$^2$ explains the calculated E2 properties.
$^{32}$Ne is more deformed; Q$_0$$\approx$65~e~fm$^2$. The calculated 2$^+$ excitation energy,  0.67~MeV  fits well
with the experimental value, 0.72~MeV. The  4$^+$ is predicted  at 1.89~MeV. Finally, for $^{31}$Ne the calculation
produces a  $\frac{3}{2}^-$ ground state which is mainly 2p-2h, and the first excited state at $\approx$200~keV  is a $\frac{3}{2}^+$ 
of 3p-3h character, belonging to the K=$\frac{1}{2}^+$  in parallel with what happens in its isotone  $^{33}$Mg. 
The intrinsic quadrupole moment of the K=$\frac{3}{2}^-$ band is  Q$_0$$\approx$60~e~fm$^2$,  and, as we mentioned before,
 the occupation of the $p$ orbits exceeds 1.2 neutrons.

 \medskip
 \noindent
{\bf  $^{29}$F  and $^{31}$F.}
\medskip

Not very much is known experimentally about $^{29}$F and $^{31}$F. Our calculations
 produce a  $\frac{5}{2}^+$ ground state  in  $^{29}$F, which is 60\%  0$\hbar \omega$  with  a first excited $\frac{1}{2}^+$ at
 0.91~MeV, which is  80\% intruder and the head of a K=$\frac{1}{2}^+$ band, as expected from quasi-SU3 and Nilsson diagrams.
 This compares fairly well with a recent  measure at Riken  \cite{riken.ca54}  which places this state  at 
 1.06~MeV.  In our calculation, the ground state of   $^{31}$F  is  an extremely mixed  $\frac{5}{2}^+$  (66\% intruder) and
 the excited $\frac{1}{2}^+$ (74\% intruder) appears at much lower excitation energy, 0.21~MeV.  Neutron excitations result in
 binding energy gains of 1.9~MeV and 2.5~MeV respectively, which may help to explain the far off location of the Fluorine's
 neutron drip line.

 \medskip
 \noindent
{\bf  $^{33}$Al  and $^{35}$Al.}
\medskip

  $^{33}$Al has its ground states largely dominated by the "normal" configurations ($\sim$80\%). Thus, it does not belong properly to the
 island of inversion.  The calculations reproduce very well the properties of the $\frac{5}{2}^+$  ground state; 
 the magnetic moment is  +4.088(5)~$\mu_N$ $vs$  +4.17~$\mu_N$(b)  and +3.86~$\mu_N$(e); the spectroscopic quadrupole moment  
 is +0.12~eb compared to the calculated +0.12~eb. Our results do not produce a low lying $\frac{5}{2}^+$ as surmised in the experiment of
 ref.~\cite{mittig}.  This is consistent with the large excitation energy of the intruder 0$^+$ in $^{34}$Si.
 The lowest excited state is  predicted to be a $\frac{5}{2}^+$ of intruder nature at 1.70~MeV 
  followed by  two other 2p-2h states at 1.85~MeV  ($\frac{1}{2}^+$) and 2.28~MeV  ( $\frac{5}{2}^+$).  Contrary to some compiled results
  we do not produce negative parity states in this range of energies.  In  the $\frac{5}{2}^+$ ground state of $^{35}$Al the "normal" configurations    
   still lead,  but barely so at 52\%. The lowest excited states,  $\frac{1}{2}^+$ at 0.63~MeV and    $\frac{3}{2}^+$ at 0.80~MeV are intruders.  
   
    \medskip
\noindent
{\bf  The limits of the "Big Island of Deformation".}
\medskip

 We have argued  already that the previously established 
 N=20 and N=28 islands of inversion/deformation merge in the Neon, Sodium and Magnesium isotopic chains, creating a bigger one (BID). 
 Referring only to the ground states, their N=19 isotopes seem to belong to it as well, and their  N=18 ones not  (but see below).
 N=31  and the neutron drip line define the west shore of the BID.
Some heavy  Aluminums, Silicons (N$\ge$26),
 Phosphors and Sulfurs  (N$\ge$28) do belong to the N=28 sector of the BID  as well, but their less neutron rich isotopes
 do not  belong to the N=20 sector.   Fluorines are  transitional, as we have just discussed.  Of 
 the N=18 isotones, only $^{28}$Ne and  $^{29}$Na can pretend to pertain to it  with 50\% and 40\%  of intruder components in their
 ground states.  For the other isotopes,  although the ground states are "normal",  quite often intruder states show up
 at  low excitation energy.

\section{Conclusions}

 We have shown that the model space comprising the $sd$-shell for the protons and the $sd-pf$ shell for the neutrons,
 together with the effective interaction {\sc sdpf-u-mix},  make it possible to describe a very large region of nuclei, in
 particular the very neutron rich nuclei at or around the N=20 "islands of inversion".  In many cases the  inversion of
 configurations produces deformed ground state bands, thus the use of the term "islands of deformation" as well.
 According to our calculations (and also to
 the meagre experimental data available), the two islands merge in the Magnesium chain.  In the calculations they  also merge in the
 Neon and Sodium chains. However this could only be checked experimentally
  if their neutron drip lines happened to be  close enough to N=28. We have studied in detail the 
 mechanisms that lead to the inversion of normal and intruder configurations, paying particular attention to the properties
 of the states at fixed  np-nh configurations and to their correlation energy gains.  We have compared the calculations to some
 selected experimental results. The ubiquitous  deformed bands  have intrinsic (electric) quadrupole moments in the
 range  Q$_0$=60-80~e~fm$^2$.  We leave for the future a full scan of the region with the {\sc sdpf-u-mix} interaction, as well as
 the study of the one and two neutron separation energies, which requires some extra monopole work,  probably including
  three body forces.

{\bf Acknowledgements.} Partially supported by the MICINN (Spain) 
                       (FPA2011-29854); by the IN2P3(France) and MICINN(Spain) (AIC11-D-648);
                       and by the Comunidad de Madrid (Spain) (HEPHACOS S2009-ESP-1473). AP
                       is  supported by  MINECOÕs (Spain) ÒCentro de Excelencia Severo OchoaÓ Programme 
                       under grant SEV-2012-0249.

\appendix

\section{The  {\sc sdpf-u-mix}  interaction}
The {\sc sdpf-u-si} interaction was designed for 0$\hbar\omega$ calculations of very neutron rich $sd$ nuclei around N=28 in a valence space comprising the full $sd$($pf$)-shell for the protons(neutrons), {\it i. e.} this interaction was defined (implicitly) with a core of $^{28}$O. Its single particle energies (SPE's) and monopoles (neutron-proton $sd$-$pf$ and neutron-neutron $pf$-$pf$) were fixed by the spectra of $^{35}$Si,
 $^{41}$Ca, $^{47}$K and $^{49}$Ca. In order to allow for the mixing among different np-nh neutron configurations across N=20, it is necessary to add to {\sc sdpf-u-si} the following new ingredients: a) The off-diagonal cross shell $sd$-$pf$ matrix elements, which we take from the Lee-Kahana-Scott G-matrix \cite{LKS} scaled as in ref.~\cite{cmnp:07}; b) The SPE's on a core of $^{16}$O: for the the $sd$-shell orbits we use always the  {\sc usd} values \cite{USD}, while for the $pf$-shell orbits we have no experimental guidance at all. Nonetheless, for any particular set  of  $pf$-shell SPE's, the neutron-neutron  $sd$-$pf$ monopoles must be chosen such as to reproduce the spectrum of
 $^{35}$Si and the N=20 gap. As the solution is not unique, we have anchored our choice to the energy of the first excited 0$^+$ state 
 in $^{30}$Mg,  because this guarantees that in our isotopic course toward N=20 the descent of the intruder states proceeds with the correct slope. 
 Indeed, at  0$\hbar\omega$ {\sc sdpf-u-mix} and {\sc sdpf-u-si} produce identical results; c)  We have incorporated to the isovector pairing of the
   $sd$-shell
the same modifications introduced in ref.~\cite{otsu}.
The pairing reduction is needed  to avoid double counting  when core excitations are
taken explicitly into account.
A recent study of the effects of  the change of the reference valence space  in the  calculation of the renormalizations of the $sd$ and $pf$ matrix elements, made in 
ref.~\cite{kamila},  gives
a robust foundation to these modifications.

  The extra bonus of the calculations
 including core excitations is that  {\sc sdpf-u-mix}  is able to give an unified description of
  the isotopic chains with Z$\le$14 and Z$>$14.  The very large span in neutron number
 that this interaction has to cope with,  brings in some global monopole problems which could be associated to three body effects. We
 have not yet completed this part of the task that would make it possible to obtain predictions for the neutron separation energies. What we have found necessary is to make a three body like reduction of the global T=1 $pf$-shell monopole to get a smooth connexion between N=20 and N=28. 
 Therefore the present version of {\sc sdpf-u-mix}  may not be the final one, although the results shown in this paper should
 not change appreciably  with its possible future evolutions.

\end{document}